\documentclass[journal=jpccck,manuscript=article,layout=traditional]{achemso}
\usepackage{graphicx}

\title{Double Perovskite Structure Induced by Co Addition to PbTiO$_3$ : Insights from DFT
and Experimental Solid State NMR Spectroscopy}

\author{Ersen Mete}\email{emete@balikesir.edu.tr}
\affiliation{Department of Physics, Bal{\i}kesir University, Bal{\i}kesir
10145, Turkey}

\author{Selda Odaba\c{s}{\i}}
\affiliation{Department of Chemical Engineering, Middle East
Technical University, Ankara 06800, Turkey}

\author{Haiyan Mao}
\affiliation{College of Material Science \& Engineering,
Nanjing Forestry University, 159 Longran Road Naijing 210037, China}
\alsoaffiliation{Chemical and Biomolecular Engineering, University of California
Berkeley, CA 94720, USA}

\author{Tiffany Chung}
\affiliation{Chemical and Biomolecular Engineering, University of California
Berkeley, CA 94720, USA}

\author{\c{S}inasi Ellialt{\i}o\u{g}lu}
\affiliation{Basic Sciences, TED University, Kolej, Ankara 06420, Turkey}

\author{Jeffrey A. Reimer}
\affiliation{Chemical and Biomolecular Engineering, University of California
Berkeley, CA 94720, USA}

\author{O\u{g}uz G\"{u}lseren}
\affiliation{Department of Physics, Bilkent University,
Ankara 06800, Turkey}

\author{Deniz Uner}\email{uner@metu.edu.tr}
\affiliation{Department of Chemical Engineering, Middle East
Technical University, Ankara 06800, Turkey}

\begin{document}
\begin{abstract}
The effects of Co addition on the chemical and electronic structure of
PbTiO$_3$ were explored both by theory and through experiment. Cobalt was
incorporated to PbTiO$_3$ during sol gel process. The XRD data of the
compounds confirmed the perovskite structure for the pure samples. The XRD
lines broadened and showed emerging cubic-like features as the Co incorporation
increased. The changes in the XRD pattern were interpreted as double perovskite
structure formation. $^{207}$Pb NMR measurements revealed a growing isotropic
component in the presence of Co. In line with the experiments, DFT calculated
chemical-shift values corroborate isotropic coordination of Pb suggesting the
formation of cubic Pb$_2$CoTiO$_6$ domains in the prepared samples.
The state-of-the-art hybrid functional first-principles calculations
indicate formation of Pb$_2$CoTiO$_6$ with cubic structure and confirms that Co
addition can decrease oxygen binding energy significantly. Experimental UV-Vis
spectroscopy results indicate that upon addition of Co, the band gap is shifted
towards visible wavelengths which was confirmed by the energy bands
and absorption spectra calculations. The oxygen binding energies were determined
by temperature programmed reduction (TPR) measurements. Upon addition of Co, TPR
lines shifted to lower temperatures and new features appeared in the TPR
patterns. This shift was interpreted as weakening of oxygen cobalt bond
strength. The change in the electronic structure by the alterations of oxygen
vacancy formation energy and bond lengths upon Co insertion are determined by
DFT calculations.

\end{abstract}


\section{Introduction}

The PbTiO$_3$ perovskite family of materials have a diverse range of
applications.\cite{Suntivich2011,Chen2013} Their
chemical and electronic structure coupled with the tunability of the band gap
and polarizability make these materials
attractive.\cite{Liao2019,Spaldin2019,Zhang2015,Erhart2014} The ease with
which they create oxygen vacancies not only diversifies their electronic
properties, but also makes them attractive chemical compounds triggering redox
reactions\cite{Eichel2011}. Furthermore, the band gap of these compounds can
also be easily tuned by doping, resulting in tunable features such as band
gaps.\cite{Beck1998,Eichel2007,Zhou2015,Zhou2016} For example, chemically doped
PbTiO$_3$ structures are utilized for hydrogen production by
photocatalytic\cite{Li2017} and photoelectrochemical\cite{Ahn2018} cells as
anode materials. Our selection of PbTiO$_3$ mainly stems from two reasons. First, 
PbTiO$_3$ is itself a good visible light photocatalyst, thus acting as an oxidizer, and its 
structural and electronic properties are well known. Second, together with the oxides of 
Co, PbO$_x$ can exchange all of its oxygen at relatively low 
temperatures.\cite{Uner2005} These properties become important when considering 
the storage of solar energy in the chemical bonds. Widespread availability of such 
technologies is crucial for off-grid localized energy production including fuel cells as well 
as for space applications. Perovskites emerge as potentially promising 
candidates\cite{Deml2014} based on a thermodynamic analysis of oxygen vacancy 
formation.\cite{Ermanoski2013,Meredig2009}

We report a fundamental study about Co added PbTiO$_3$ (PCTO) materials
with particular attention to solid-state $^{207}$Pb NMR spectroscopy, where the
effect of Co is reflected in the chemical shifts and the lineshapes of the
signal. Magnetic resonance methods are versatile tools for simultaneously
elucidating the geometric structure along with the electronic
properties\cite{Bykov2014,Zhao1999}. Density functional theory (DFT)
is used\cite{Zeigler1999, Alkan2015} to determine the source of the NMR shifts
as well as the changes in the electronic structures and in the UV--Vis spectra
of Co-doped PbTiO$_3$.

\section{Methodology}

\noindent\textbf{Preparation of the Materials~} \\
Lead (II) Acetate Trihydrate (LAT) (extra pure, MERCK), titanium (IV)
isopropoxide (97\%, Sigma-Aldrich) (TIP), Cobalt (II) Acetate Tetrahydrate
(pure, MERCK) (CAT) were used as precursors and citric acid (99\%, Aldrich)
(CA) as gelation agent. LAT was dissolved in the minimum amount of acetic acid
glacial and 50\% by volume mixture of ethanol-acetic and acid glacial was used
to dissolve TIP and ethanol is used to dissolve citric acid. The molar ratio of
LAT, TIP and CA was (1:1:2), respectively. The solution containing TIP was
added to the solution containing LAT under vigorously stirring. As soon as
CA-ethanol solution was introduced to the mixture, a white gel was formed. The
gel was kept under vigorous stirring and later it was transferred on the hot
plate. A white solidified gel was obtained as a result. Afterwards, the white
solidified gel was placed in the oven and kept there overnight at 100 $^\circ$C
to remove volatile organic solvents and water. The white powder was heated to
650 $^\circ$C at a rate of 5 $^\circ$C/min for calcination and kept at
that temperature for 3 hours. After calcination, a yellow powder was obtained.
The same procedure was applied to prepare cobalt containing samples with the
following modifications. Appropriate amount of Cobalt (II) Acetate Tetrahydrate
(pure, MERCK) (CAT) was dissolved in the LAT-acetic acid solution. The molar
ratios of LAT, CAT, TIP and CA were adjusted to attain the desired $x=$ Co/(Pb+Co)
to prepare PCTO. After calcination, a gray powder was obtained for $x=$ 0.875, a
green powder was obtained for $x=$ 0.75 and a green powder was obtained for $x=$ 0.5.

\vspace{5mm}\noindent\textbf{Characterization~} \\
XRD analysis was used to determine phase identification and the crystallinity of
PbTiO$_3$ and PCTO samples. Analyses were performed in Philips PW 1840
Compact X-ray Diffractometer equipment (-30kV, 24mA- with Cu K$\alpha$
radiation). The scattering angle was from 5 to 90$^\circ$.
The compositions of synthesized perovskites are analyzed by using Perkin Elmer
Optima 4300DV inductively coupled plasma--optical emission
spectrometry.

Solid-State NMR Spectroscopy $^{207}$Pb static SS-NMR spectra were collected on
a 11.7 T magnet at a $^{207}$Pb frequency of 104.53 MHz. A Bruker narrow bore
H/C/N probe was used. A $^{207}$Pb 90$^\circ$ pulse of 3.3 s was measured in
solid Pb(NO$_3$)$_2$. Pb(NO$_3$)$_2$ was used as a secondary chemical shift
reference with the left horn of powder pattern was set to 3490 ppm (relative
to Me$_4$Pb). $^{207}$Pb Hahn echo experiments were performed with an
interpulse delay of 20 s. The relaxation delay was used as 60 s for
PbTiO$_3$, 30 s for PCTO samples. All the NMR measurements were
obtained at room temperature ($\approx$ 25 $^\circ$C).

UV-Vis spectroscopy analysis was done in Shimadzu UV-2450 equipment. The
absorbance data was measured between 200 nm and 800 nm. Barium sulfate was 
used as reference sample.

Micromeritics Chemisorb 2720 equipment was used for TP$_\textrm{x}$ analyses.
Temperature programmed reduction (TPR), temperature programmed thermal
decomposition (TPtD), temperature programmed oxidation (TPO) photodesorption
experiments were conducted to measure the reducibility of samples before and
after UV and heat treatments. The composition of the effluent gases are tracked
by a thermal conductivity detector (TCD). Before letting the gas flow through
the system, the samples are placed between quartz wool in the U-shaped quartz
reactor and the quartz reactor is placed in a furnace which can be heated up to
1100 $^\circ$C. A cold trap, a mixture of ice, water, and isopropyl alcohol is
used to remove condensables, particularly water vapor from the product stream
before the analysis. The final temperature, heating rate and stand by time at
final temperature are fixed using TP$_\textrm{x}$ controller. UV-desorption
experiments were performed under a flow of He while maintaining a UV
irradiation over the sample.

\vspace{5mm}\noindent\textbf{Computational Details~} \\
Density functional calculations have been performed with the Vienna
ab-initio simulation package (VASP)\cite{Kresse1993} using the
projector-augmented wave (PAW) method\cite{Blochl1994,Kresse1999}.
The single particle states have been expanded in plane waves up to
a kinetic energy cutoff value of 400 eV. The exchange and correlation
effects were taken into account using the modern hybrid
Heyd--Scuseria--Ernzerhof (HSE)\cite{HSE03,HSE06,Paier2006} scheme.

The standard density functionals are known to be insufficient for describing
the perovskite materials\cite{Umeno2006}. Estimation of their structural and
vibrational properties improves with meta-GGA
functionals\cite{Paul2017,Zhang2017}. In the absence of a proper
self-interaction cancellation between the Hartree and exchange terms leads to a
significant band gap underestimation. Hybrid approaches have been proposed to
improve the description of electronic structures over usual GGA
functionals.\cite{Bilc2008} The screened Coulomb hybrid density functional,
HSE\cite{HSE03,HSE06,Paier2006}, partially incorporates the exact Fock exchange
and the PBE\cite{PBE1996} exchange energies. The
HSE\cite{HSE03,HSE06,Paier2006} correlation energy and the long-range (LR) part
of the exchange energy is taken from the PBE\cite{PBE1996} functional. The
short-range (SR) part of the exchange energy is mixed with the PBE counterpart
using $\eta$ as the mixing coefficient~\cite{Perdew1996} as,
\[
E_{\textrm{\tiny X}}^{\textrm{\scriptsize HSE}}=
\eta E_{\textrm{\tiny X}}^{\textrm{\scriptsize HF,SR}}(\omega)+
(1-\eta)E_{\textrm{\tiny X}}^{\textrm{\scriptsize PBE,SR}}(\omega)+
E_{\textrm{\tiny X}}^{\textrm{\scriptsize PBE,LR}}(\omega)
\]
where $\omega$ is the range separation parameter.\cite{HSE03,HSE06,Paier2006}
We employed the HSE12s\cite{Moussa2012} functional which optimizes these
parameters to reduce the Fock exchange length scale without decreasing the overall
accuracy of HSE06\cite{HSE06} significantly. This range separated hybrid
density functional approach improves the band gap related properties over the
standard exchange--correlation (XC) schemes and offers a better description of
localized $d$ states of transition metals. In particular, the position and
dispersion of possible Co-driven gap states are important for Co-doped
PbTiO$_3$. Recently, the hybrid DFT approach has been successfully employed
to get the electronic structures of perovskite
oxides.\cite{Weston2016a,Weston2016b}

In order to determine the structure of the perovskite with Co, a
(2$\times$2$\times$2) supercell was constructed from the bulk unit cell of
PbTiO$_3$. Then we traced possible interstitial and substitutional Co incorporation
models with various Co/(Pb+Co) ratios. Geometry optimization of these Co incorporated 
supercells ended up with tetragonal symmetry. The Brillouin zone integrations have been
carried out over a $\Gamma$-centered 8$\times$8$\times$8 $k$-point grid.
Both the cell volume and the atomic positions were fully optimized
self-consistently until the Hellmann--Feynman forces on each ion in each
cartesian direction was less than 0.01 eV/{\AA}.

The DFT--NMR calculations were performed based on the linear response GIPAW
method.\cite{Pickard2001,Yates2007,Gregor1999} The chemical shift tensor
component values at a nuclear site at \textbf{R} were determined from,
\[
 \delta_{ij}(\textrm{\bf R})=\frac{\partial B_i^{\textrm{\tiny
ind}}(\textrm{\bf R})}{\partial B_j^{\textrm{\tiny ext}}}
\]
where $i$ and $j$ are the cartesian indices, $B^{\textrm{\tiny
ext}}$ is an applied DC external magnetic field, and $B^{\textrm{\tiny
ind}}(\textrm{\bf R}$) is the induced magnetic field at \textbf{R}.
Then a proper referencing to TML  is done using
\[ \delta_{\mathrm{iso}}^{\mathrm{calc.}}(^{207}\mathrm{Pb}) =
\delta_{\mathrm{iso}}^{\mathrm{ref}} -
\delta_{\mathrm{iso}}^{\mathrm{PCTO}}
~~~~ \mathrm{(all \: in \: ppm)},    \]
\noindent where a single TML molecule is considered
in a large computational cell.

Technically, NMR chemical shift calculations require a higher cutoff value
than the usual. We set it to 600 eV. A more stringent tolerance is also needed
to stop the self-consistent loop. We used a value of $10^{-10}$ eV for the
difference of the total energies between the electronic iterations. \\

\section{Results \& Discussion}

\subsection{Structure Analysis}

\begin{figure}[h!]
\centering
\includegraphics[width=8.5cm]{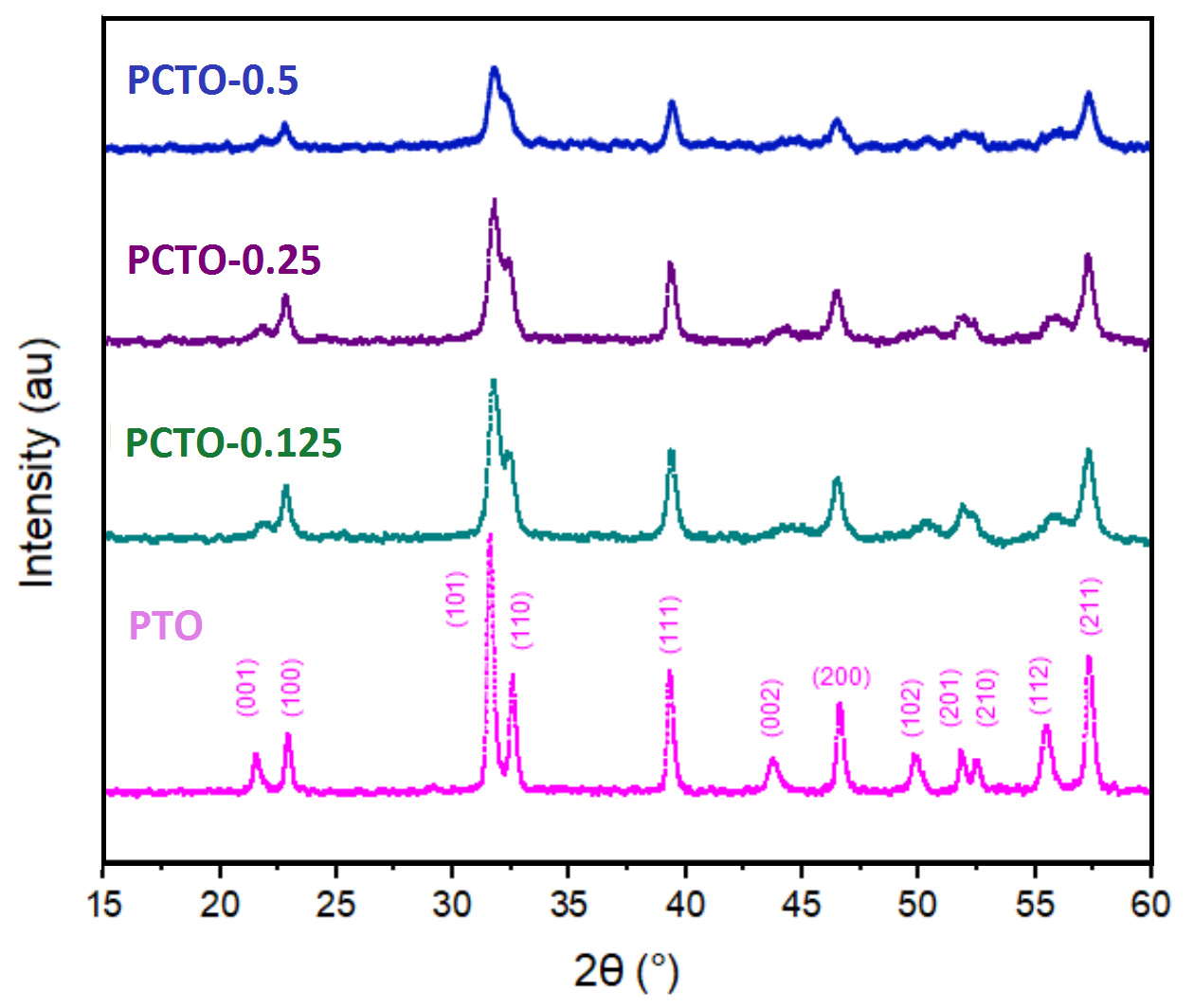}
\caption{(color on-line) XRD patterns of PbTiO$_3$ and PCTO with
various Co/(Pb+Co) ratios ($x=$ 0.125, 0.25, and 0.5).\label{fig_xrd}}
\end{figure}

Structure of the synthesized samples of PTO and PCTO with Co/(Pb+Co) ratios
$x=$ 0.125, 0.25,and 0.5 were determined using XRD analysis 
(See Fig.~\ref{fig_xrd}). Upon Co addition, the peaks corresponding to (101) and 
(110), (201) and (210), (112) and (211) broaden and merge. As Co content 
increases in the samples, cubic planes become dominant to the detriment of 
tetragonality. In addition, as cobalt content increased, intensity of the peaks 
decreased, suggesting the decrease of the crystal grain size as cobalt content 
increased.

\begin{figure}[h!]
\centering
\includegraphics[width=16.2cm]{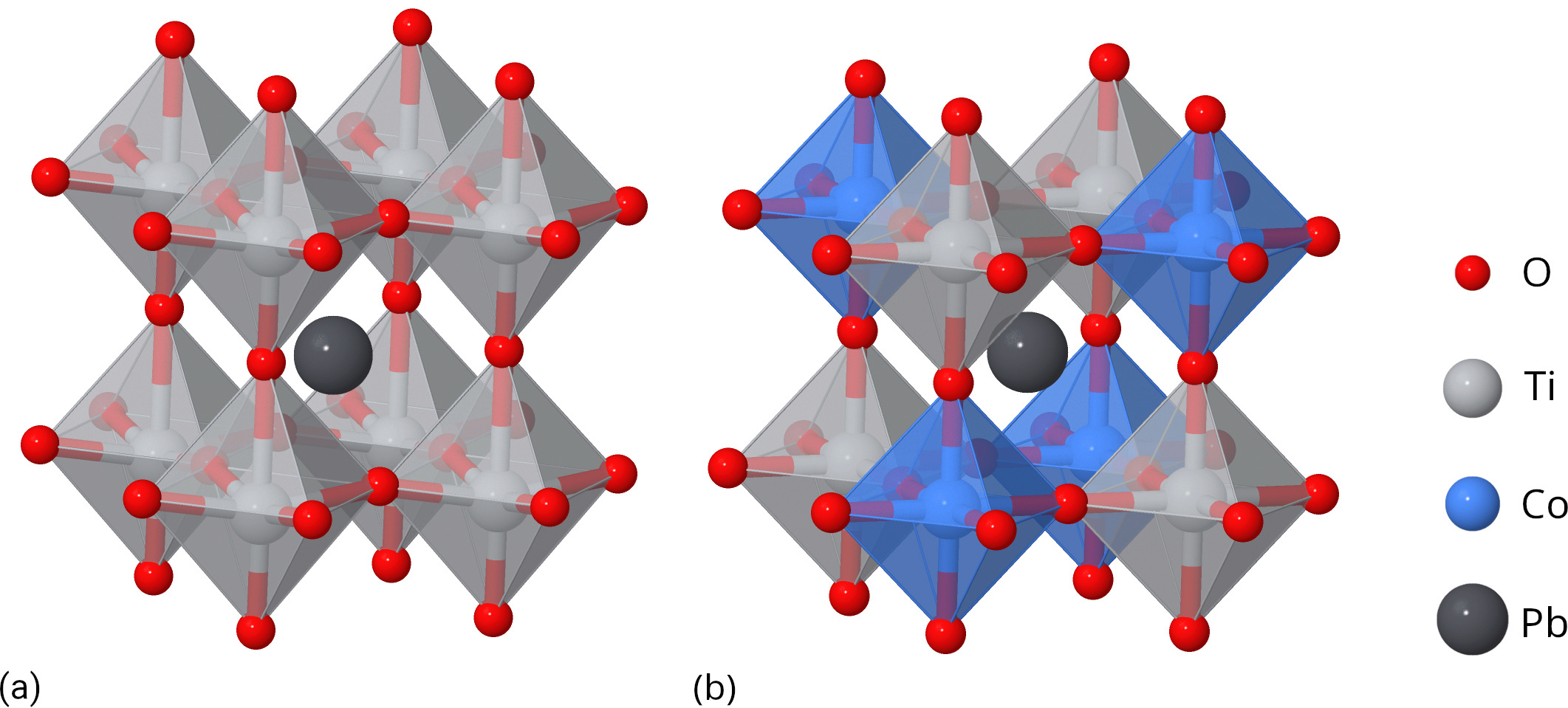}
\caption{(color on-line) Schematic (a) tetragonal PbTiO$_3$ and (b) cubic
Pb$_2$CoTiO$_6$ structures.\label{fig_struct}}
\end{figure}

\begin{table*}[h!]
\small
\caption{Atomic positions of tetragonal ($P$4$mm$) PbTiO$_3$ and cubic
($Fm\bar{3}m$) Pb$_2$CoTiO$_6$.  DFT results were obtained using the HSE
XC-functional.\label{table1}}
\begin{tabular}{cccccccccccccc}\hline\\[-1.8mm]
\multicolumn{8}{c}{PbTiO$_3$} && \multicolumn{5}{c}{Pb$_2$CoTiO$_6$} \\[2mm]
\hline\\[-1.8mm]
Atom & x$^*$ & y$^*$ & z$^*$ && x & y & z && Atom & x & y & z & Site \\[1mm]
\hline\\[-1.8mm]
Pb & 0.000 & 0.000 & 0.000 && 0.000 & 0.000 & 0.000 && Pb & 1/4 & 1/4 & 1/4
& 8c\\[1.2mm]
Ti & 0.500 & 0.500 & 0.530 && 0.500 & 0.500 & 0.535 && Co & 1/2 & 1/2 & 1/2 &
4b\\[1.2mm]
O1 & 0.500 & 0.500 & 0.074 && 0.500 & 0.500 & 0.087 && Ti & 0 & 0 & 0 &
4a\\[1.2mm]
O2 & 0.500 & 0.000 & 0.641 && 0.500 & 0.000 & 0.610 && O & 0.2563(4) & 0.000 &
0.000 & 24e\\[1mm] \hline
\end{tabular}
\begin{flushleft}
$^*$ \footnotesize{Experimental values for PbTiO$_3$ are taken from
Ref.\cite{Sahu2011}~.}
\end{flushleft}
\end{table*}

The tetragonal phase of PbTiO$_3$ is shown in Fig.~\ref{fig_struct}a.
Since, the experimental XRD data indicate the presence of a cubic phase in
the Co added PbTiO$_3$. We checked all possible cubic structures including
substitutional (for Pb and/or Ti) and interstitial dopings. The only consistent
cubic structure obtained is for Pb$_2$CoTiO$_6$ as shown in
Fig.~\ref{fig_struct}b.

We performed geometry optimization calculations without imposing space
group symmetry and allowed a full volume relaxation by lifting any
restriction on the lattice translation vectors. At 325 K, PbTiO$_3$ forms the
tetragonal phase having $P$4$mm$ space group symmetry with cell parameters
$a=3.899$ {\AA} and $c=4.138$ {\AA}.\cite{Noheda2000} The lattice constants
were found as $a=3.864$ {\AA} and $c=4.045$ {\AA} using the HSE12s functional. 
The largest deviation from the experimental values comes from the $c$ parameter 
which is $\sim$2.2$\%$. This is known as the super-tetragonality problem of HSE 
functional which is based on local density approximation (LDA). A recent theoretical 
study reported similar values with reasonable agreement using the HSE06 XC 
functional.\cite{Zhang2017} The atomic coordinates are  presented in Table~\ref{table1}.

The change in the XRD peaks as the Co content increases in the samples
indicates the presence of a cubic phase in relation to Co. In addition, SS-NMR
measurements reveal isotropic chemical environment for Pb atoms in
the PCTO materials. Preparation process of Co added PbTiO$_3$ leads to
formation of Pb$_2$CoTiO$_6$ domains which involve TiO$_6$ and CoO$_6$
octahedra as shown in Fig.~\ref{fig_struct}b. The double perovskite has a
cubic $Fm\bar{3}m$ symmetry ($a=7.68$ {\AA}) where Co and Ti sit at the corners
of the cube in an alternating manner. In this structure, a Pb atom, being
at the center, is 12-fold-coordinated with the nearest oxygens which lie at the
mid-points of the edges of the cube.

\subsection{NMR Spectroscopy}

\begin{figure}[h!]
\includegraphics[width=0.5\textwidth]{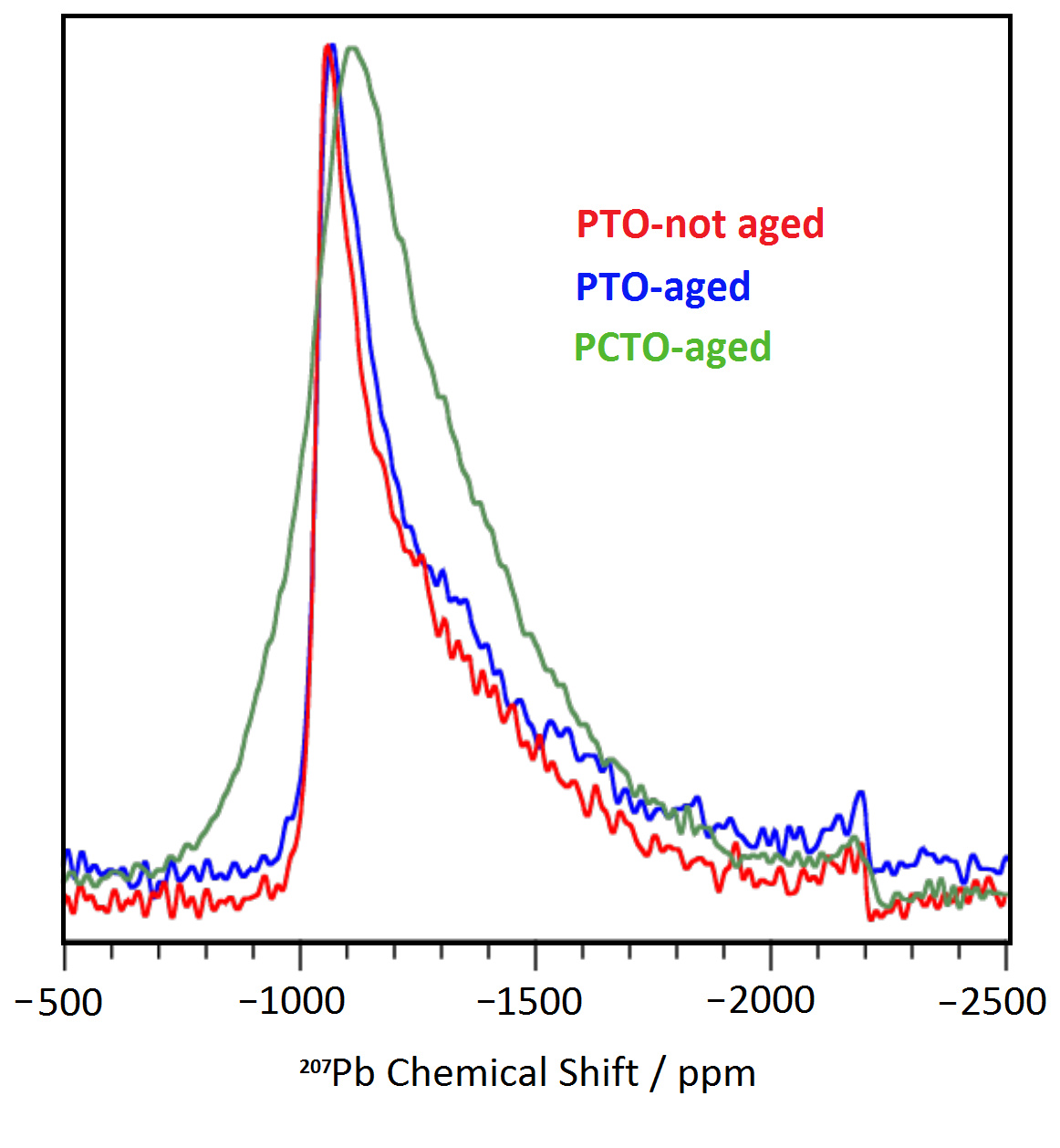}
\caption{(color on-line) The effect of aging during sol gel and addition of a
trace amount of Co (PCTO) on the NMR lineshape of $^{207}$PbTiO$_3$.
\label{fig_nmr1}}
\end{figure}

Hahn echo data from pure PbTiO$_3$ (red, not aged), PbTiO$_3$ (blue, aged one
month at 280 K) and PbTiO$_3$ with trace amount of cobalt, also aged one month
at 280 K (green) are shown in Fig.~\ref{fig_nmr1}. NMR spectra of the pure compounds
are consistent with the reports from the literature~\cite{vanBramer} while cobalt doped
sample reveals a broadening at the left horn, and 100 ppm shift to lower frequencies.

\begin{figure}[h!]
\includegraphics[width=0.5\textwidth]{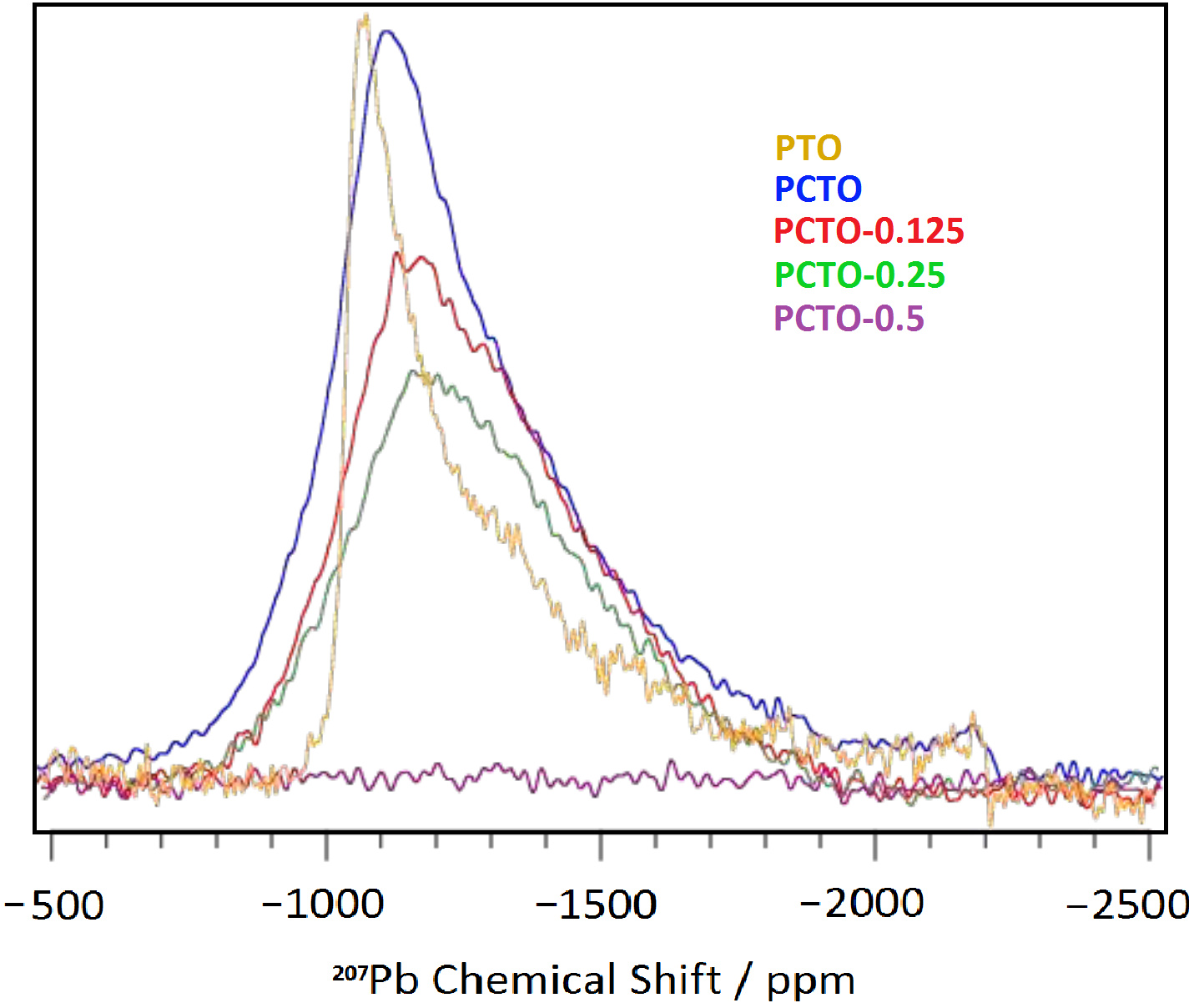}
\caption{(color on-line) The effect of Co addition on the NMR lineshape of
$^{207}$PbTiO$_3$. \label{fig_nmr2}}
\end{figure}

The NMR spectra of Co added PbTiO$_3$ are shown in Fig.~\ref{fig_nmr2}.
When high amounts of cobalt was introduced in the sample at the percent
levels to replace Pb and/or Ti atoms in the perovskite lattice, the NMR lineshape
evolved into a broad, featureless form centered around the horn of the pure
PbTiO$_3$ spectrum. At the intended 50\% replacement of the Pb atoms with Co,
the intensity diminished completely, both due to broadening and due to the
diminished amount of Pb nuclei in the sample. The total intensities of the other
Co containing samples were consistent with the intensity of the pure compound,
indicating that most of the nuclei were accessible by NMR.

\begin{figure}[h!]
\includegraphics[width=0.8\textwidth]{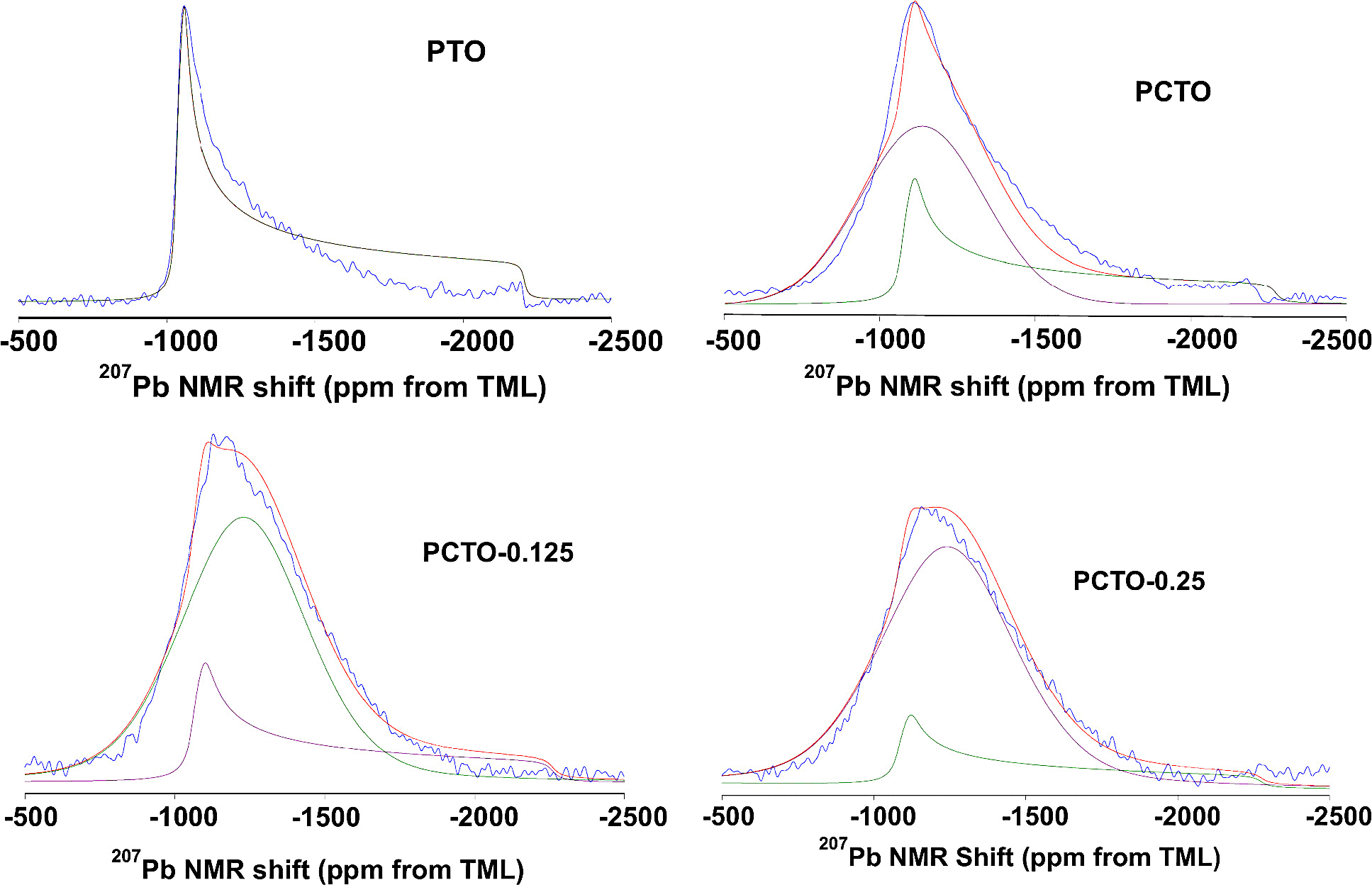}
\caption{(color on-line) NMR lineshape simulations.\label{fig_nmr3}}
\end{figure}

NMR linesphape simulations shown in Fig.~\ref{fig_nmr3} reveal superposition of
two features. One feature is the unperturbed PbTiO$_3$ lineshape, the second
feature is a broad symmetric isotropic peak. These $^{207}$Pb NMR
data suggest coexistence of Pb related two different local phases in the
samples. The chemical shift anisotropy (CSA) tensor component estimations of
lineshape simulations are consolidated in Table~\ref{table2}

\begin{table}[h!]
\small
\caption{DM Fitting, Chemical shift anisotropy+Gaus/Lor \label{table2}}
\begin{tabular}{lccccccc}\hline\\[-1.8mm]
Compound  & $\delta_{\footnotesize\textrm{xx}}$ &
$\delta_{\footnotesize\textrm{yy}}$ &
$\delta_{\footnotesize\textrm{zz}}$ & $\delta_{\footnotesize\textrm{iso}}$
& $\delta_{\footnotesize\textrm{aniso}}$ &
$\eta$ & Span\\[1mm] \hline\\[-1.8mm]
PTO &  $-1032.50$ & $-1059.59$ & $-2207.02$ & $-1433.04$ & $-773.98$ & 0.0063 & 1174.52
\\[1mm] \hline\\[-1.8mm]
PCTO-trace & $-1052.88$ & $-1076.48$ & $-2244.76$ & $-1458.04$ & $-786.72$ & 0.0054 &
1191.87 \\[1.2mm]
PCTO-0.125 & $-1124.20$ & $-1165.90$ & $-2270.52$ & $-1520.21$ & $-750.33$ & 0.0091 &
1120.59 \\[1.2mm]
PCTO-0.25 & $-1138.05$ & $-1150.75$ & $-2252.80$ & $-1513.84$ & $-738.88$ & 0.0028 &
1101.05 \\[1.2mm] \hline
\end{tabular}
\end{table}

\begin{table*}[h!]
\small
\caption{Chemical shift anisotropy values (in ppm) estimated from DFT in
tetragonal PbTiO$_3$ and in cubic Pb$_2$CoTiO$_6$.
\label{table3}}
\begin{tabular}{lccc}\hline\\[-1.8mm]
model structure  &$\delta_{\footnotesize\textrm{xx}}$ &
$\delta_{\footnotesize\textrm{yy}}$ & $\delta_{\footnotesize\textrm{zz}}$
\\[1mm] \hline\\[-1.8mm]
PbTiO$_3$ &  $-$1032 & $-$1032 & $-$1538 \\[1mm] \hline\\[-1.8mm]
Pb$_2$CoTiO$_6$ & $-$1148 & $-$1148 & $-$1148 \\[1.2mm]
\end{tabular}
\end{table*}

This analysis of NMR spectra suggests an isotropic structure, so we searched
scenarios thoroughly. We considered all possible models for Co inclusion to
PbTiO$_3$ lattice such as substitutional (for Pb and/or Ti) and interstitial
dopings. The only probable geometry is found as Pb$_2$CoTiO$_6$ which forms in
a cubic structure where oxygen and Pb coordination is isotropic.

The DFT calculations are used to elucidate the role of Co dopants on the
chemical shift of $^{207}$Pb. The local chemical environments are reflected
better in the diagonal components of chemical shift anisotropy tensor estimated
from DFT are compiled in Table~\ref{table3}. The left horn of the $^{207}$Pb
NMR spectrum in PbTiO$_3$ is a result of symmetry characteristics of Pb--O
bonding (mainly two types of Pb--O bonds with lengths 2.75 {\AA} and 3.09 {\AA})
in pure tetragonal PbTiO$_3$. Co is mixed at the sample preparation stage.
Therefore, oxygen octahedra forms with Co and/or Ti at the center which leads
to formation of a cubic Pb$_2$CoTiO$_6$ double perovskite structure. In this
structure Pb--O coordination is isotropic. Therefore, Pb--O bonds become similar
in length and in covalency. Since the valence electron distribution around each
of the Pb nucleus gets affected by Co incorporation, their shielding responses
to an external magnetic field show the isotropic nature this cubic structure.
Although the relativistic effects are not included in the GIPAW implementation
of VASP, the DFT calculations are successful in reproducing $^{207}$Pb NMR
features reasonably close to experimental spectra.

\subsection{Oxygen Bonds}

The effect of a single oxygen vacancy formation on the PbTiO$_3$ lattice
structure is considered using a (2$\times$2$\times$2) supercell containing 40
atoms. In the absence of an oxygen atom both from the PbO layer (O1) and from
the TiO$_2$ layer (O2), small lattice distortions occur and remain in the local
environment. The oxygen vacancy formation energy can be formulated as
\[
E_\textrm{f}=E_{\textrm{\tiny PCTO}}-E_{\textrm{\tiny PCTO}'}-E_{\textrm{\tiny O}}
\]
where $E_{\textrm{\tiny PCTO}}$ and $E_{\textrm{\tiny PCTO}'}$
are the total cell energies of PbTiO$_3$ without and with an
oxygen defect, respectively. $E_{\textrm{\tiny O}}$ is the energy of an oxygen
atom in the O$_2$ molecule for which the calculation procedure is adopted from
Ref.~\cite{Unal2014} The energy required to remove one oxygen atom
from the undoped bulk system is as large as 5.28 eV per O1 and 5.14 eV per O2.
In Pb$_2$CoTiO$_6$ structure, the chemical bonding characteristics of all oxygens
are equal and the vacancy formation energy is 3.12 eV. Therefore, Co addition to
PbTiO$_3$ leads to a significant drop in the oxygen vacancy formation energy.

\begin{figure}[h!]
\centering
\includegraphics[width=8cm]{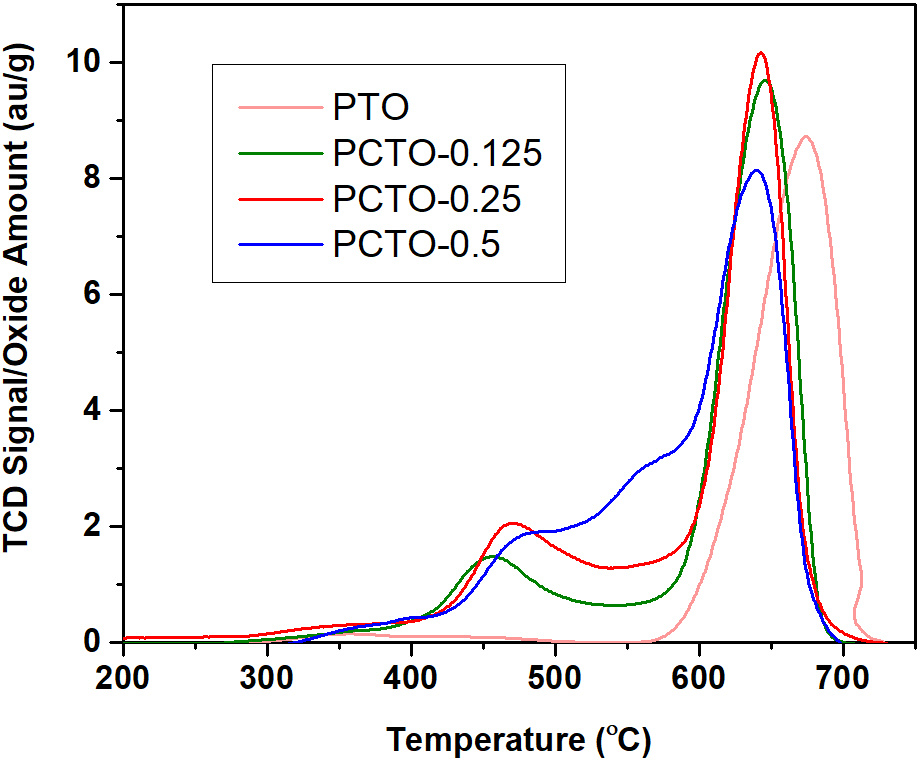}
\caption{(color on-line) TPR profiles of perovskites.\label{fig_tcd}}
\end{figure}

\begin{table}[h!]
\small
\caption{TPR peak positions of synthesized perovskites \label{table4}}
\begin{tabular}{lcl}\hline\\[-1.8mm]
Sample  & Peak Positions ($^\circ$C) & Additional Peaks ($^\circ$C) \\[1mm]
\hline\\[-1.8mm]
PTO & 675 & --\\[1mm]
PCTO-0.125 & 646 & 455 \\[1.2mm]
PCTO-0.25 & 643 & 158, 469  \\[1.2mm]
PCTO-0.5 & 640 & 400, 478, 558  \\[1.2mm] \hline
\end{tabular}
\end{table}

Temperature programmed reduction (TPR) profiles of the perovskites are
consolidated in Fig.~\ref{fig_tcd}. It is clearly seen that the addition of Co
decreased the reduction temperature of the material, indicative of weaker
oxygen bonds in these structures.  The peak temperatures were compiled in
Table~\ref{table4} for comparison. Here, the 675 $^\circ$C peak can be observed
for all perovskites. As Co is introduced, peak position
shifts to 646 $^\circ$C  which indicates a weaker oxygen bond energy
in the structure. Further addition of Co does not significantly influence the
peak position.

\begin{figure}[h!]
\includegraphics[width=\textwidth]{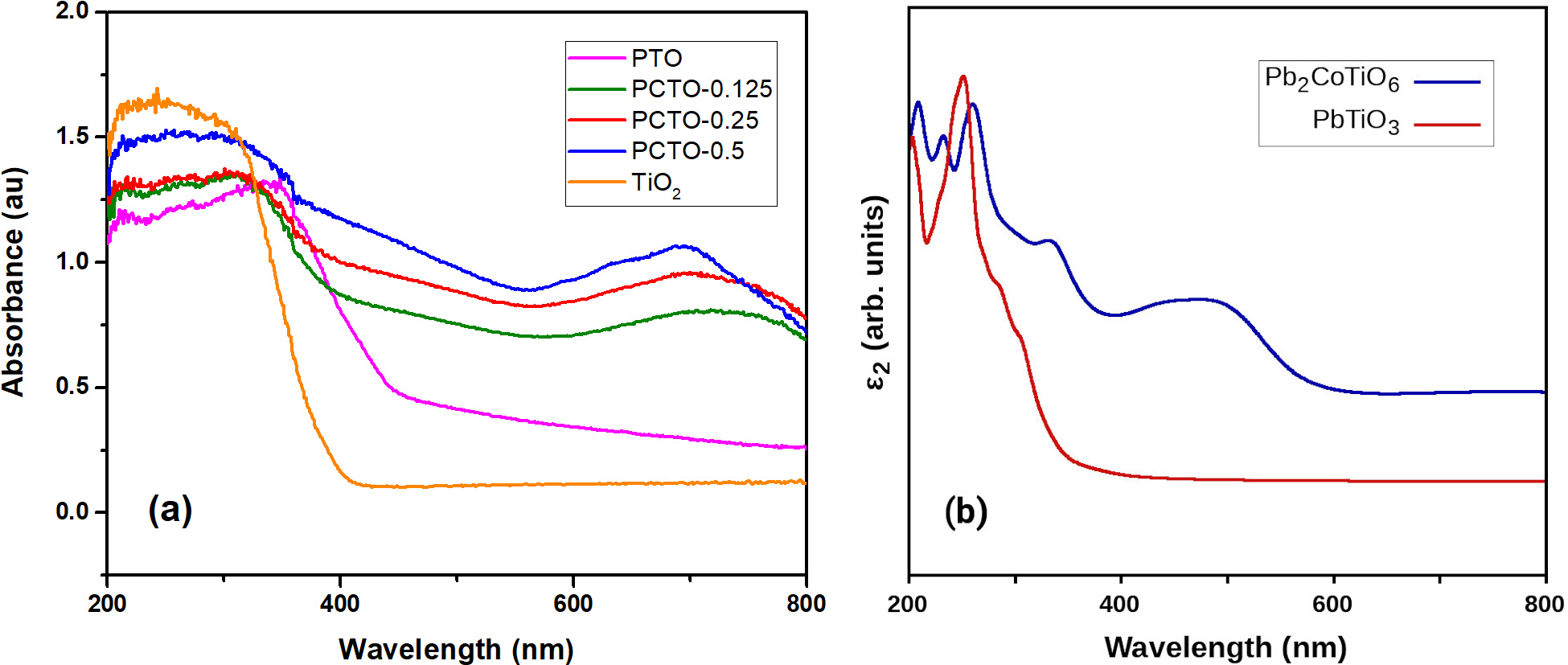}
\caption{(color on-line) (a) UV-Visible Diffuse Reflectance Spectra of TiO$_2$,
pure and Co added PbTiO$_3$. (b) HSE-calculated imaginary part of the
dielectric
matrix of PbTiO$_3$ and
Pb$_2$CoTiO$_6$. \label{fig_optics}}
\end{figure}

\subsection{Electronic structure}

In Fig.~\ref{fig_optics}a, experimentally measured UV-Visible diffuse
reflectance spectra is shown. All synthesized perovskites and TiO$_2$ had
strong absorption between 200 nm--400 nm. For PbTiO$_3$, absorption edge is
larger than the absorption edge of TiO$_2$. The result conforms with the
literature~\cite{Choi2010,Deng2014}. The peak around 650 nm--800 nm for
Co-doped samples was originated from Co$^{4+}$ species. Dai \textit{et al.}
showed that Co$_3$O$_4$ and Co$_3$O$_4$/TiO$_2$ had similar absorption spectra
between 600 nm--800 nm while for pure TiO$_2$, there are no peaks in this
region\cite{Dai2013}. Band gap of TiO$_2$ and PbTiO$_3$ was estimated as 3.1 eV
and 2.5 eV, respectively. Similar, to the $^{207}$Pb NMR results, increasing Co
content did not further change the characteristics of the experimental UV-Vis
spectra.

The imaginary part of the dielectric matrix $\epsilon_2$ as a function of the
wavelength was calculated for PbTiO$_3$ and Pb$_2$CoTiO$_6$ and presented in
Fig.~\ref{fig_optics}b. The effect of Co addition to PTO on the main features
of absorption spectra agrees with the experimental results. The
agreement between the experimental and theory predicted spectra further
confirms the double perovskite structure.

\begin{figure*}[h!]
\centering
\includegraphics[width=16.3cm]{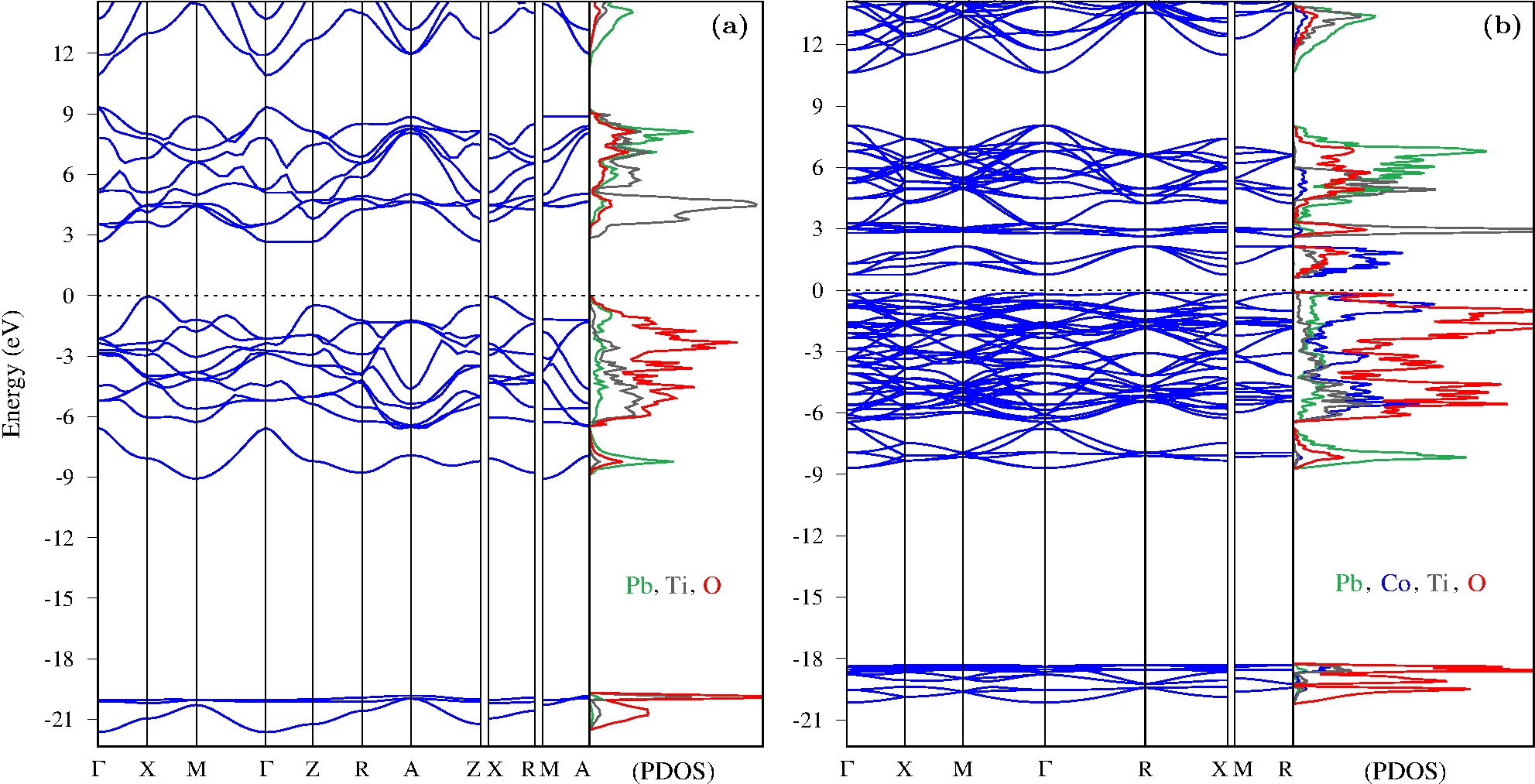}
\caption{(color on-line) Electronic energy band structures and corresponding
projected densities of states (PDOS) of (a) tetragonal PbTiO$_3$ and (b) cubic
Pb$_2$CoTiO$_6$ calculated using the HSE12s DFT functionals. The zero of the
energy eigenvalues are referenced to the Fermi energy which is depicted as
dotted lines.\label{fig_bands}}
\end{figure*}

Electronic energy bands and the corresponding partial densities of states (PDOS)
are calculated using the HSE functional, for tetragonal  PbTiO$_3$ and simple
cubic Pb$_2$CoTiO$_6$ (See Fig.~\ref{fig_bands}). At first glance, the main
features common to perovskites can be seen in both of the materials such as the
compositions of core levels, VB,  and CB. First of all, the computational method
makes a difference in the results. The HSE functional brings a large correction in
the band gaps over the standard DFT functionals. For instance, the indirect band gap
(X$\Gamma$) of PbTiO$_3$ predicted by using the PBE functional is about 1.5 eV
and the width of the valence band (VB) is approximately 8.5 eV. The HSE
functional, however, leads to a gap about 2.8 eV in much better agreement with the
experiment (2.5 eV) and the VB width is close to 9 eV in
Fig.~\ref{fig_bands}a. Pb$_2$CoTiO$_6$ is a semiconductor with an indirect band
gap (R$\Gamma$)  of 0.88 eV. The main reason of this drastic difference between
the band gaps of  pure and Co added PbTiO$_3$ is due to the Co 3\emph{d} driven
antibonding  $t_{2g}$--orbitals. These states appear 0.47 eV below the CB as a satellite
group which has a width of 1.4 eV.

The bottom of the CB at about 3 eV in Fig.~\ref{fig_bands}b is characterized
dominantly by well localized Ti 3\emph{d} $e_g$-type orbitals. Antibonding states
associated with Pb mostly contribute to the upper part of the CB.

The single band in Fig.~\ref{fig_bands}a starting from 0.3 eV below the valence
group has a width of 2.6 eV and consists mostly of Pb-6\emph{s} character.
Similarly, in the case of PCTO, a group of bands separated from the
bottom of the VB group have the same Pb-6\emph{s} character. The
core level, mostly of  O-2\emph{s} character, is at $-20.2$ to $-21.8$ eV in
(a), whereas the same bands in (b) are shifted 1.9 eV upward and range from
$-18.3$ to $-20.6$ eV.

\section{Conclusions}

The effects of cobalt incorporation into PbTiO$_3$ perovskites are investigated
experimentally as well as by the first principles methods. Consistent with the
common expectations, Co replaces lattice Ti. The insertion of cobalt transforms
tetragonal PbTiO$_3$ to a cubic Pb$_2$CoTiO$_6$ confirmed by XRD, NMR and UV-Vis
data as well as their corresponding DFT simulations. The DFT estimated values
of $^{207}$Pb NMR chemical shift anisotropy tensor components could predict the
symmetry around different nuclei with the changes in Co content.  NMR lineshape
simulations reveal an isotropic structure in addition to the characteristic
PbTiO$_3$ lineshape. Based on the DFT investigations, the most probable geometry
is found as Pb$_2$CoTiO$_6$ which forms in a cubic structure where oxygen and Pb
coordination is isotropic. The oxygen vacancy formation energy decreases
significantly upon addition of Co as estimated by DFT in line with the TPR
measurements. Co incorporation in the PbTiO$_3$ structure causes a significant
narrowing of the band gap as a result of the Co related $t_{2g}$-type $d$-states
($\sigma*$) which appear near the bottom of the conduction band.

\section{Acknowledgements}
The authors acknowledge the contributions from Dr. Taymaz Tabari and 
Dr. Jun Xu at the initial phase of this study. Financial support for 
the experimental part of this study was provided by T\"{U}B\.{I}TAK under 
grant no 107M040.

\bibliography{refs}

\end{document}